%% file: main.tex
\documentclass[preprint,12pt]{elsarticle}

\usepackage{amssymb}
\usepackage{amsmath}

\input{macro.tex}

\journal{Computer Methods and Programs in Biomedicine}

\begin{document}

\begin{frontmatter}



\title{An Integrated Open Source Software System for the Generation and Analysis of
Subject-Specific Blood Flow Simulation Ensembles}


\author[label1]{Simon Leistikow\fnref{equal1}}
\author[label1]{Thomas Miro\corref{cor1}\fnref{equal1}}
\author[label2]{Adrian Kummerländer}
\author[label3]{Ali Nahardani}
\author[label4]{Katja Grün}
\author[label5,label6]{Markus Franz}
\author[label3]{Verena Hoerr}
\author[label2]{Mathias J. Krause}
\author[label1]{Lars Linsen}

\cortext[cor1]{Corresponding author. Email: tmiro@uni-muenster.de}
\fntext[equal1]{These authors contributed equally to this work.}

\affiliation[label1]{organization={Computer Science Department, University of Muenster},
            addressline={Einsteinstraße 62},
            city={Münster},
            postcode={48149},
            state={North Rhine-Westphalia},
            country={Germany}}

\affiliation[label2]{organization={Institute of Applied and Numerical Mathematics, Karlsruhe Institute of Technology},
            addressline={Englerstraße 2},
            city={Karlsruhe},
            postcode={76131},
            state={Baden-Württemberg},
            country={Germany}}

\affiliation[label3]{organization={Heart Center Bonn, Department of Internal Medicine II, University Hospital},
            addressline={Venusberg-Campus 1},
            city={Bonn},
            postcode={53127},
            state={North Rhine-Westphalia},
            country={Germany}}

\affiliation[label4]{organization={Department of Internal Medicine I, University Hospital Jena},
            addressline={Am Klinikum 1},
            city={Jena},
            postcode={07747},
            state={Thuringia},
            country={Germany}}

\affiliation[label5]{organization={Department of Cardiology, Angiology and Intensive Care Medicine, Cardiovascular Center Hersfeld-Rotenburg},
            addressline={Heinz-Meise-Straße 100},
            city={Rotenburg an der Fulda},
            postcode={36199},
            state={Hesse},
            country={Germany}}

\affiliation[label6]{organization={Department of Cardiothoracic Surgery, University Hospital Jena},
            addressline={Am Klinikum 1},
            city={Jena},
            postcode={07747},
            state={Thuringia},
            country={Germany}}

\begin{abstract}
\textbf{Background and Objective:} 
Hemodynamic analysis of blood flow through arteries and veins is critical for diagnosing cardiovascular diseases, such as aneurysms and stenoses, and for investigating cardiovascular parameters, such as turbulence and wall shear stress. For subject-specific analyses, the anatomy and blood flow of the subject can be captured non-invasively using structural and 4D Magnetic Resonance Imaging (MRI). Computational Fluid Dynamics (CFD), on the other hand, can be used to generate blood flow simulations by solving the Navier-Stokes equations. To generate and analyze subject-specific blood flow simulations, MRI and CFD have to be brought together. \\[1mm]
\textbf{Methods:} 
We present an interactive, customizable, and user-oriented visual analysis tool that assists researchers in both medicine and numerical analysis. Our open-source tool is applicable to domains such as CFD and MRI, and it facilitates the analysis of simulation results and medical data, especially in hemodynamic studies. It enables the creation of simulation ensembles with a high variety of parameters. Furthermore, it allows for the visual and analytical examination of simulations and measurements through 2D embeddings of the similarity space. \\[1mm]
\textbf{Results:} 
To demonstrate the effectiveness of our tool, we applied it to three real-world use cases, showcasing its ability to configure simulation ensembles and analyse blood flow dynamics. We evaluated our example cases together with MRI and CFD experts to further enhance features and increase the usability.\\[1mm]
\textbf{Conclusions:} 
By combining the strengths of both CFD and MRI, our tool provides a more comprehensive understanding of hemodynamic parameters, facilitating more accurate analysis of hemodynamic biomarkers.
\end{abstract}


\end{frontmatter}


\section{Introduction}\label{sec:introduction}
Numerical simulations are essential for understanding complex phenomena across scientific disciplines, including medicine. In hemodynamics, computational fluid simulations help investigate the blood flow within blood vessels. Configuring these simulations can be tedious, especially when trying to assimilate measured data, as setting boundary conditions and ensuring stability demands a careful setup. Moreover, medical imaging data must first be transformed into 3D surface models, typically requiring multiple separate software tools.

Integrating these steps into a single application can improve productivity and flexibility. Yet, existing commercial and open-source solutions often target specific use cases with limited adaptability, restricting their applicability. In \autoref{sec:related work}, we compare state-of-the-art software tools for (medical) flow simulation and analysis, highlight their strengths and weaknesses and point out a resulting gap in medical simulation research.

In particular, generating (subject-specific) \textbf{simulation ensembles} remains cumbersome, and integrated support for \textbf{analyzing} such data is often missing. Researchers frequently resort to custom scripts, limiting reproducibility and generalizability. This highlights the need for a unified application that efficiently combines specialized tools for flow simulation and analysis.

We address need this by integrating the Lattice Boltzmann simulation framework OpenLB~\cite{krause2021openlb} into Voreen~\cite{drees2022voreen}, our visualization and analysis framework.Our tool combines the high-performance simulation capabilities of OpenLB with Voreen’s customizable workflows. An overview of both frameworks is given in \autoref{sec:methods} after the background in \autoref{sec:background}.

To demonstrate the tool’s capabilities, we define research use cases from computational fluid dynamics (CFD) and cardiovascular imaging, developed in collaboration with domain experts. Based on these, we build user-friendly and customizable workflows in Voreen to address the use cases (\autoref{sec:results}) and evaluate them through expert feedback (\autoref{sec:discussion}).

The individual contributions of our work can be summarized as follows:
\begin{itemize}
    \item We present an extensible, customizable, and integrated system with a user-friendly graphical user interface (GUI) to configure, perform, and analyze three-dimensional Lattice Boltzmann simulations. 
    \item We demonstrate the effectiveness of our system in three research-oriented use cases: \\[1mm]    
    (A) The configuration of simulations based on a given geometry stored in a stereolithography (STL) file.\\[1mm]
    (B) The preparation and configuration of subject-specific simulations based on medical imaging data. \\[1mm]
    (C) The visual analysis of the subject-specific simulation ensemble created in (B).
    \item We evaluate our software with feedback from researchers from the CFD and cardiovascular imaging domain, respectively.
\end{itemize}

\section{Review of Existing Software Solutions}\label{sec:related work}
Our tool for the generation of simulation ensembles and their visual data analysis relates to research in computational fluid dynamics, cardiovascular imaging, and visualization. While these fields have specific challenges, they share methods like generating and visualizing velocity vector fields. Consequently, established frameworks face both domain-specific and overlapping issues.

Numerous frameworks and comparative studies exist, e.g. \cite{CFD_comparison, cfd_comparison_winter2014benchmark}. Voreen, for instance, has been compared to similar tools \cite{drees2022voreen}, but only regarding visualization.
We review existing software solutions for fluid simulations and their analysis. Since the range of available tools is broad, we focus on the most comparable in scope and functionality to our system. To ensure objectivity, we define categories for evaluation. Key factors for existing solutions are \textbf{availability}, \textbf{usability}, and \textbf{supported features}.   
Availability covers licensing model (open source vs. commercial) and platform compatibility. Usability includes customization options, required expertise (e.g., CFD or MRI knowledge), documentation, and user support.  
Comparing supported features is more complex. For our purposes, relevant aspects include 3D model generation from volumetric medical imaging data, simulation and ensemble creation, analysis, and postprocessing functionalities.
\autoref{tab:comparison_factors} summarizes these categories and criteria.

\begin{table}[htb]
\centering
\begin{tabular}{|l|l|}
\hline
\textbf{Category} & \textbf{Comparison Factors} \\ \hline
Availability & Licensing model (open source vs. commercial) \\ \cline{2-2}
             & Supported operating systems \\ \hline
Usability & Required prior knowledge (e.g., CFD, MRI) \\ \cline{2-2}
          & Customization options \\ \cline{2-2}
          & Availability of documentation \\ \cline{2-2}
          & User support (e.g., community forums) \\ \hline
Supported Features & 3D model generation from imaging data (Preprocessing) \\ \cline{2-2}
                   & Simulation and ensemble creation \\ \cline{2-2}
                   & Simulation and ensemble analysis \\ \cline{2-2}
                   & Graphical User Interface (GUI)  \\ \cline{2-2}
                   & Interactive Visual Data Analysis (Postprocessing) \\ \hline
\end{tabular}
\caption{Comparison categories and factors for evaluating biomedical simulation software.}
\label{tab:comparison_factors}
\end{table}


\medskip
\textit{SimVascular} \cite{updegrove2017simvascular, SimVascular_webpage} is an open-source software designed as a complete pipeline for blood flow modeling, from medical image segmentation to hemodynamic simulations. It offers functionalities for preprocessing, including image segmentation, discrete solid modeling, mesh repair tools, and fluid-solid interaction modeling with variable wall properties. Additionally, it provides a highly configurable simulation setup, supported by well-structured documentation, making it accessible for detailed hemodynamic studies.  
However, visualization and analysis are outsourced to \textit{ParaView}, making multi-simulation analysis tedious due to manual export/import across tools.

\textit{CRIMSON} \cite{arthurs2021crimson,crimson_webpage} is an open-source software designed as an advanced simulation environment for subject-specific hemodynamic analysis. Developed as a descendant of \textit{SimVascular}, it shares a common foundation, particularly in medical image segmentation capabilities. However, its development has emphasized computational hemodynamics (CH), incorporating a wide range of adjustable boundary conditions to enhance simulation accuracy. Additionally, unlike SimVascular, CRIMSON integrates visualization and postprocessing within its own graphical user interface, which is available on Linux and Windows.  
Despite its strengths, CRIMSON targets individual simulations and lacks built-in support for parallel ensemble analysis, limiting its use for parameter studies and sensitivity analyses.

\textit{SimScale} \cite{SimScaleWebsite, SimScaleDocs} is a cloud-based simulation platform designed for a range of engineering applications. It is a commercial, web-based solution that provides a user-friendly interface. In the context of (multiphase) flow simulations, SimScale allows users to perform geometric modeling, configure flow properties, and define boundary conditions. Additionally, it offers postprocessing tools for visualizing individual simulations. The platform enhances usability through tutorials, and professional customer support.  
However, it lacks support for simulation ensembles with varying parameters and built-in comparative analysis of multiple simulations on the same object. Additionally, it does not preprocess medical imaging data (DICOM, NIfTI). While external tools can be used to convert medical data into CAD-compatible formats, this requires additional expertise, making the workflow less accessible to users from the medical domain.

Comparable commercial software such as \textit{Ansys} \cite{AnsysWebsite} also provides fluid flow simulation capabilities. Since its advantages and limitations closely align with those of SimScale in the context of our comparison, we only mention it briefly here.

\textit{FabSim3} \cite{groen2023fabsim3, FabSim3Docs} is an automation toolkit designed for verified simulations using high-performance computing. At its core, it is a Python-based framework that allows users to write custom scripts for running simulation ensembles and analyzing their outputs with various metrics. By automating key steps in pre- and post-processing, it minimizes manual errors and reduces the workload associated with managing large-scale simulations. Additionally, it is supported by a documentation that details its functionalities.  
However, it does not include a GUI, making it less accessible to users without a background in computer science or related fields. Furthermore, while it offers a high degree of automation, its parameter configuration options are more constrained compared to other solutions, which may limit flexibility in the simulation setup. 

\textit{OpenFOAM} \cite{openfoam_webpage, chen2014openfoam, jasak2009openfoam} is an open-source C++ library and one of the most widely adopted computational mechanics software solutions. Compared to FabSim3, it provides a more modular approach, allowing users to fine-tune a wide range of settings for both pre- and post-processing. It also supports the computation of simulation ensembles with varying boundary conditions through user-written scripts. Additionally, OpenFOAM has gained significant popularity, leading to an active community of contributors and users, which ensures continuous development and improvement.  
However, as a library, OpenFOAM lacks a GUI, reducing accessibility for non-experts in computer science (or related fields). Unlike FabSim3, it requires extensive CFD knowledge due to minimal automation and manual configuration. Visualization relies on external tools like ParaView.

\textit{Voreen} \cite{drees2022voreen, voreen, voreen-github} is an open-source framework for the interactive visualization and analysis of volumetric data. It offers highly configurable analysis and visualization workflows. Within these workflows, volume processing pipelines can be built and executed. The framework supports medical image processing and visualization as well as the connection to computing clusters for expensive tasks such as flow simulations. Moreover, it provides a GUI, allowing users to visualize processing and simulation results in various ways and, thus, enhances their visual analysis. A more rigorous description is provided in \autoref{sec:methods}.
However, Voreen offers limited support for geometric modeling. Although it includes basic smoothing and hole-filling algorithms, most modeling tasks are delegated to external CAD software. Moreover, boundary conditions are auto-detected via vessel graph extraction and only slightly adjustable. Further limitations are discussed in \autoref{sec:discussion} and \autoref{sec:conclusion}.

\autoref{tab:feature-overview} summarizes supported features. Availability and usability are excluded, focusing solely on technical capabilities for simulation and visualization.
Each reviewed software has strengths and limitations for different user groups. Our solution proposed in this paper enables ensemble simulations and analysis in a user-friendly environment requiring minimal prior knowledge. Integrating all steps into one system removes the need for external software and manual data transfers, filling a gap in medical simulation research. Our system combines Voreen with OpenLB (see Section~\ref{sec:methods}) and is, therefore, labeled ``Voreen+OpenLB" in \autoref{tab:feature-overview}.

\newcommand{\fullCapable}{\tikz[baseline=-0.5ex]\filldraw[black] (0,0) circle (0.9ex);}
\newcommand{\partCapable}{\tikz[baseline=-0.5ex]{
  \begin{scope}
    \clip (0,0) circle (0.9ex);
    \fill[black] (0ex,-0.9ex) rectangle (1.1ex,0.9ex);
  \end{scope}
  \draw (0,0) circle (0.9ex);}}
\newcommand{\noneCapable}{\tikz[baseline=-0.5ex]\draw[black] (0,0) circle (0.9ex);}

\begin{table*}[ht]
    \centering
    \renewcommand{\arraystretch}{1.5} 
    \resizebox{\textwidth}{!}{%
    \begin{tabular}{@{}lcccccc@{}} 
        \toprule
        \textbf{Feature} & \textbf{Voreen+OpenLB} & \textbf{SimVascular} & \textbf{Crimson} & \textbf{SimScale} & \textbf{FabSim3} & \textbf{OpenFOAM} \\
        \midrule
        Medical Image Segmentation      & \fullCapable & \fullCapable & \fullCapable & \noneCapable & \partCapable & \noneCapable \\
        Geometric Modeling              & \partCapable & \fullCapable & \partCapable & \fullCapable & \noneCapable & \fullCapable \\
        Configure Boundary Conditions   & \partCapable & \fullCapable & \fullCapable & \fullCapable & \partCapable & \fullCapable \\
        Running Simulations             & \fullCapable & \fullCapable & \fullCapable & \fullCapable & \fullCapable & \fullCapable \\
        Running Simulation Ensembles    & \fullCapable & \fullCapable & \partCapable & \partCapable & \fullCapable & \fullCapable \\
        In-Situ Analysis of Simulations & \fullCapable & \noneCapable & \partCapable & \fullCapable & \partCapable & \partCapable \\
        Post-Processing \& Visualization& \fullCapable & \partCapable & \fullCapable & \fullCapable & \partCapable & \partCapable \\
        Open Source                     & \fullCapable & \fullCapable & \fullCapable & \noneCapable & \fullCapable & \fullCapable \\
        \bottomrule
        \multicolumn{7}{l}{\footnotesize Legend: $\fullCapable$ = fully supported, $\partCapable$ = partially supported, $\noneCapable$ = not supported} \\
    \end{tabular}
    }
    \caption{
        Overview of supported features in different simulation and visualization software solutions. The feature classification into “fully supported”, “partially supported”, and “not supported” is a simplification to aid comparison. Actual capabilities may vary depending on use cases, configurations, or ongoing developments.
    }
    \label{tab:feature-overview}
\end{table*}

\section{Background on Flow Simulation Ensembles}\label{sec:background}
Visualization, simulation, and analysis often depend on multiple parameters. Instead of examining them separately, simulation ensembles allow for analyzing combined parameter settings \cite{wilson2009toward, evers2022multi}. 
In our work, we define a flow simulation ensemble as a set of simulated time-resolved 3D flow vector fields as in \cite{hummel2013comparative}. Each individual simulation is called an ensemble member.

\subsection{Lattice Boltzmann Methods}
The macroscopic motion of fluids is commonly described using the \emph{Navier--Stokes equations} (NSE). We use the force-free incompressible NSE
\begin{align}
\begin{cases}
    \partial_t u + (u \cdot \nabla) u = -\frac{1}{\rho} \nabla p + \nu \Delta u & \text{in } \Omega \times I \\
    \nabla \cdot u = 0 & \text{in } \Omega \times I
\end{cases}\label{eq:nse}
\end{align}
for velocity \(u\), pressure \(p\), density \(\rho > 0\), and kinematic viscosity \(\nu > 0\) on spatial domain \(\Omega \subseteq \mathbb{R}^3\) and time \(I\subseteq \mathbb{R}_{>0}\).
This macroscopic model is sufficient for hemodynamic simulations, as blood can be assumed to be incompressible under all physiological conditions. 

\begin{figure}
\centering
\input{figures/d3q19.tikz}
\caption{The discrete velocity set \(D3Q19\)\cite{simonis2023pde}}
\label{fig:lattice}
\end{figure}
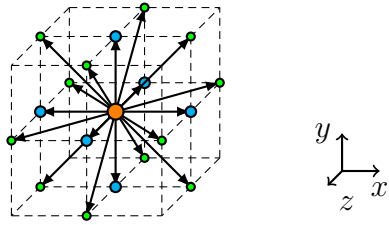

The \emph{Lattice Boltzmann method} (LBM) is used to approximate the NSE~\eqref{eq:nse} target equation on a regular space-time grid with the \(D3Q19\) velocity stencil (cf. \autoref{fig:lattice}).
More information about Lattice Boltzman methods and solutions approaches can be found in \cite{kruger,simonis2023pde}.

\section{Our Integrated Software System}\label{sec:methods}
As discussed in \autoref{sec:related work}, existing solutions address parts of the challenges in creating, visualizing, and analyzing simulation ensembles, but lack an interactive and customizable tool tailored for CFD and cardiovascular imaging experts. We address this by integrating the high-performance Lattice Boltzmann framework \textit{OpenLB} into the interactive visual analysis framework \textit{Voreen}.

\subsection{Voreen}\label{frameworks-Voreen}
\textit{Voreen} \cite{drees2022voreen, voreen, voreen-github} is a C++ open-source framework for interactive visualization and analysis of multi-modal volumetric data. It provides volume rendering, interactive data analysis, and configurable analysis workflows. It is maintained in a public repository on GitHub~\cite{voreen-github} and available for Windows and Linux. In the following, we present features relevant to our use cases in \autoref{sec:results}. A more detailed feature overview is given in \cite{drees2022voreen}.


\subsubsection*{Geometry processing and rendering}
Voreen supports the import and export of geometric file formats like Wavefront Object Files (.obj) and Stereolithography Files (.stl). Basic processing, such as hole-filling for watertight meshes, is available. Rendering options include order-independent transparency. Geometries can be voxelized via inside-outside tests.

\subsubsection*{General volume processing and rendering}
Voreen provides volume processing tools such as iso-surface extraction, connected component analysis, centerline extraction, and random walker segmentation. Scalar volume data can be visualized using, for example, slice-based, or direct volume rendering by (out-of-core) raycasting. For vector-valued volume data, streamline, pathline, and pathsurface rendering can be chosen. 

\subsection{OpenLB}\label{frameworks-openlb}

\emph{OpenLB}~\cite{krause2021openlb,olbug17,openlb18} is an open-source C++ framework for multi-physics simulations using the Lattice Boltzmann (LB) method.
It provides a broad suite of LB models enabling transparent execution on both CPU and GPU architectures~\cite{kummerlaender2023,kummerlaender2024,kummerlander_optimization_2024}. 
Beyond this, OpenLB covers workflows for pre- and post-processing. Simulation cases are fully differentiable, enabling sensitivity analysis and adjoint optimization~\cite{Krause2010,krause_parallel_2013}.

\begin{figure}[h]
    \centering
    \includegraphics[width=\linewidth]{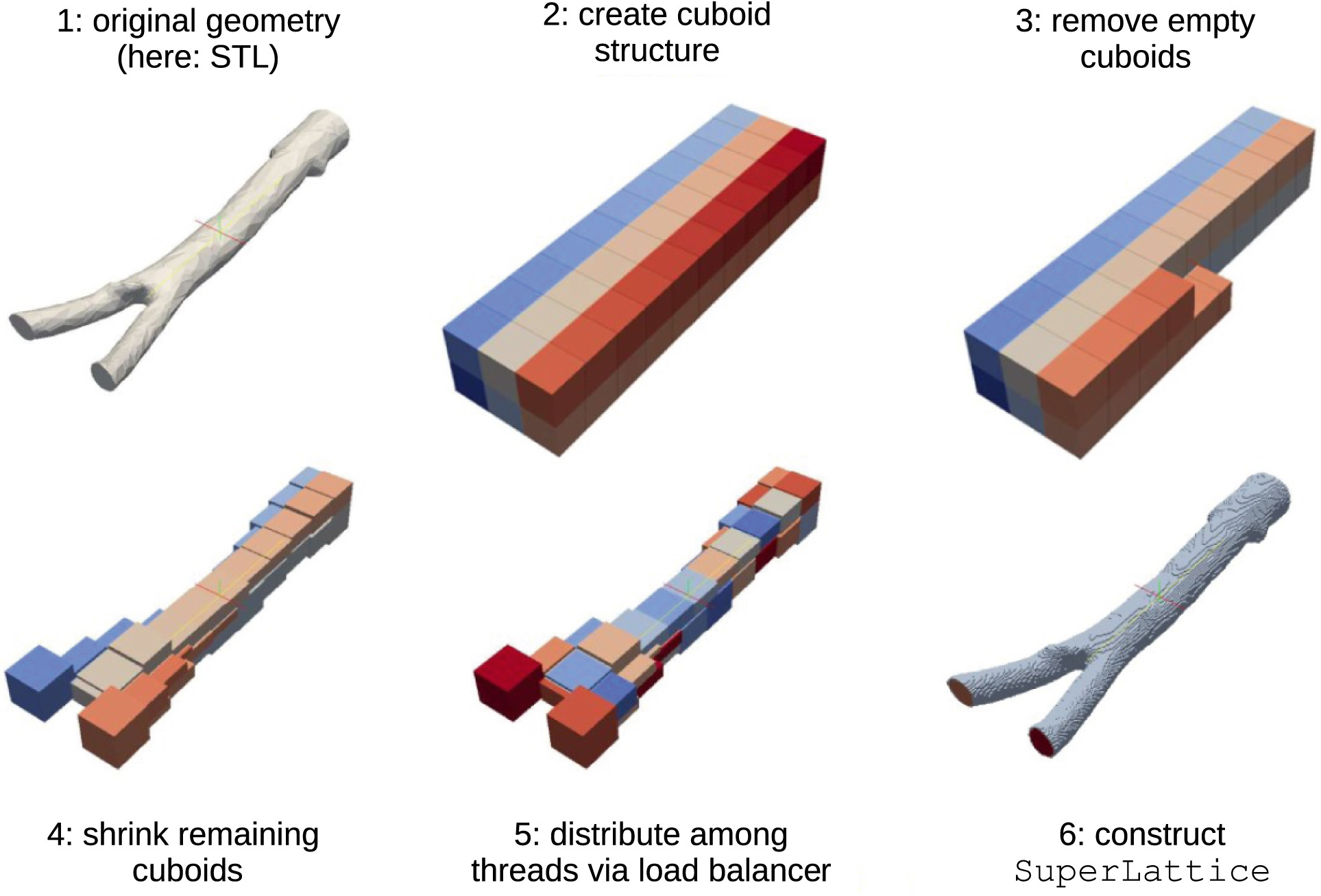}
    \caption{Automated meshing and decomposition~\cite{krause2021openlb}}
    \label{fig:preprocessing}
\end{figure}

\subsection{Integration}\label{ss:integration}

We integrate OpenLB into Voreen as a flexible \emph{backend} module, reading simulation parameters from Voreen's XML parameter file.
This case is compiled once with a suitable choice of parallelization modes and then executed via the Voreen UI. 

\begin{figure*}[ht]
    \centering
    \includegraphics[width=1.0\linewidth]{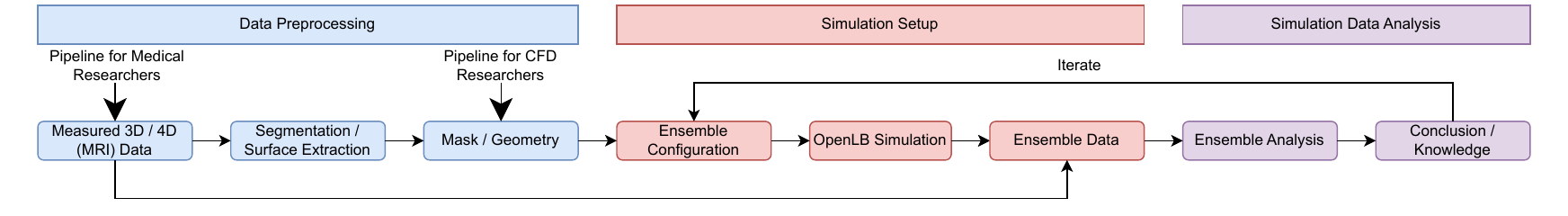}
    \caption{Ensemble Analysis Pipeline: Processing and configuration of simulation ensembles as well as their analysis.}
    \label{fig:ensemble_pipeline}
\end{figure*}

This integration extends Voreen’s feature set with fluid flow simulation capabilities.
Together, Voreen and OpenLB provide a comprehensive tool for volumetric data processing. \autoref{fig:ensemble_pipeline} illustrates workflows emphasizing Voreen's unique selling point: all pipeline steps — from segmentation to analysis — are performed within a single software, allowing iterative adjustments of any pipeline step with small effort.

\section{Results}\label{sec:results}
This section introduces three use cases for typical CFD and MRI tasks, demonstrating the setup and analysis of simulation ensembles with interactive visualization techniques \cite{leistikow2020interactive}. Unlike Leistikow et al., who emphasized technical parameters, we focus on providing a practical user guide. The use cases cover: (A) configuring ensembles from STL files, (B) generating 3D models and ensembles from MRI data, and (C) analyzing ensembles from (B) to derive insights.
Detailed guides are available in the GitHub Wiki for \href{https://github.com/voreen-project/voreen/wiki/OpenLB-UseCase-A}{OpenLB Use-Case A}, \href{https://github.com/voreen-project/voreen/wiki/OpenLB-UseCase-B}{OpenLB Use-Case B} and \href{https://github.com/voreen-project/voreen/wiki/OpenLB-UseCase-C}{OpenLB Use-Case C}. Furthermore, in the supplementary material we provide tutorial videos for all use cases.

\subsection{Use Case A: Simulation for given geometry}\label{uc: uc-a}
\begin{figure}[ht]
    \centering
    \includegraphics[width=\columnwidth]{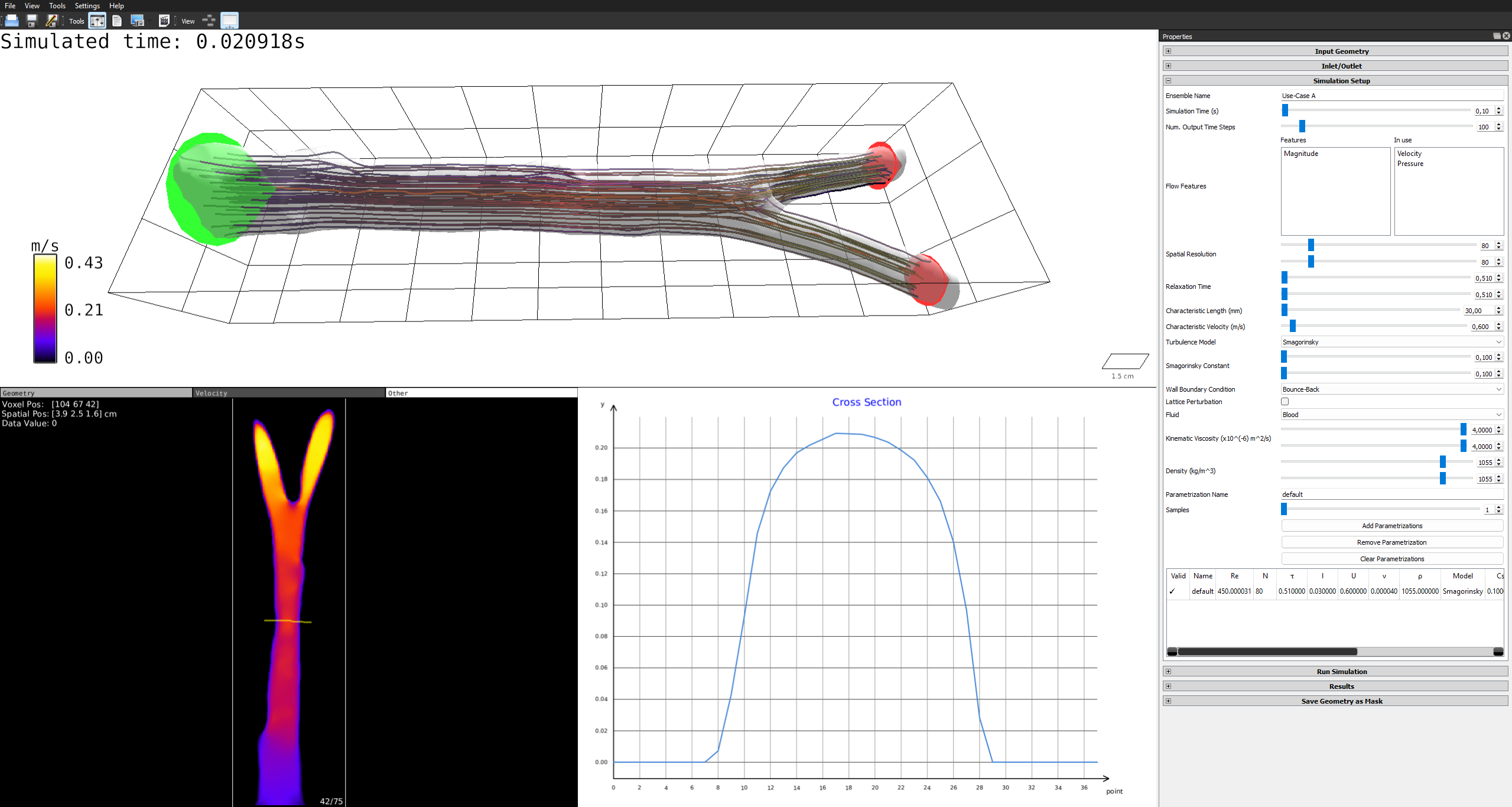}
    \caption{In-situ simulation result for Use Case A. Multiple linked views visualize the geometry and streamlines (top), axial velocity magnitude map (bottom left). The user can sketch a line in the latter view for which the velocity magnitude profile is rendered in the bottom right view. Tabs in the bottom left view allow to switch between visualizing the material volume used by OpenLB for setting boundary conditions, the velocity magnitude and an alternative output such as the pressure field.  Plots are updated in real time as the simulation is running, as soon as a new time step has been written.}
    \label{fig:uc-a}
\end{figure}

CFD researchers often work with geometry or volume files, such as blood vessels in stored in .stl format \cite{STL-explanation}). Challenges such as boundary condition setup and processing are typically handled in a trial-and-error iteration of parameter selection and result analysis. In-situ visualization during runtime can accelerate this process by enabling early aborts and reconfiguration. Use Case A implements this workflow, with user interaction reduced to the following steps:
\begin{itemize}
\item Open the predefined workspace \textbf{use\_case\_A.vws} and select a geometry file from the \textit{geometry input} panel. 
\item Configure and adjust simulation parameters from the \textit{Inlet/Outlet} and \textit{Simulation Setup} panel.
\item Initiate the calculation process by using the \textit{Run Simulation} function, and save the results in a preferred format and folder using the \textit{Results} function. 
\end{itemize}
\autoref{fig:uc-a} shows a sample result for this simulation.

\subsection{Use Case B: Subject-specific simulation for given medical imaging data }\label{uc: uc-b}
\begin{figure}[ht]
    \centering
    \includegraphics[width=\columnwidth]{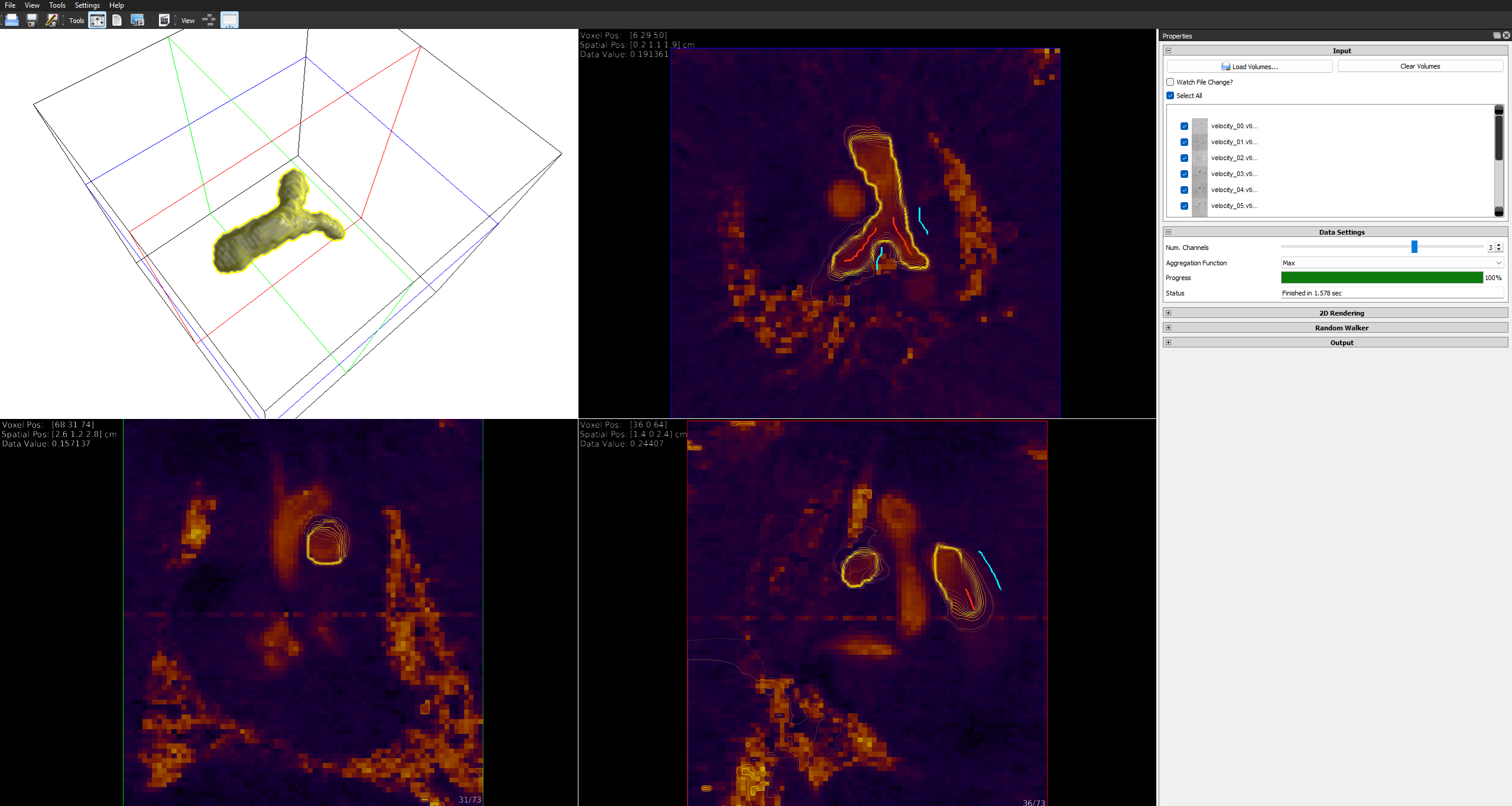}
    \caption{Segmentation result for Use Case B: upper left the segmentation, other windows display axial, coronal, and sagittal planes.}
    \label{fig:uc-b_segmentation}
\end{figure}

\begin{figure}[ht]
    \centering
    \includegraphics[width=\columnwidth]{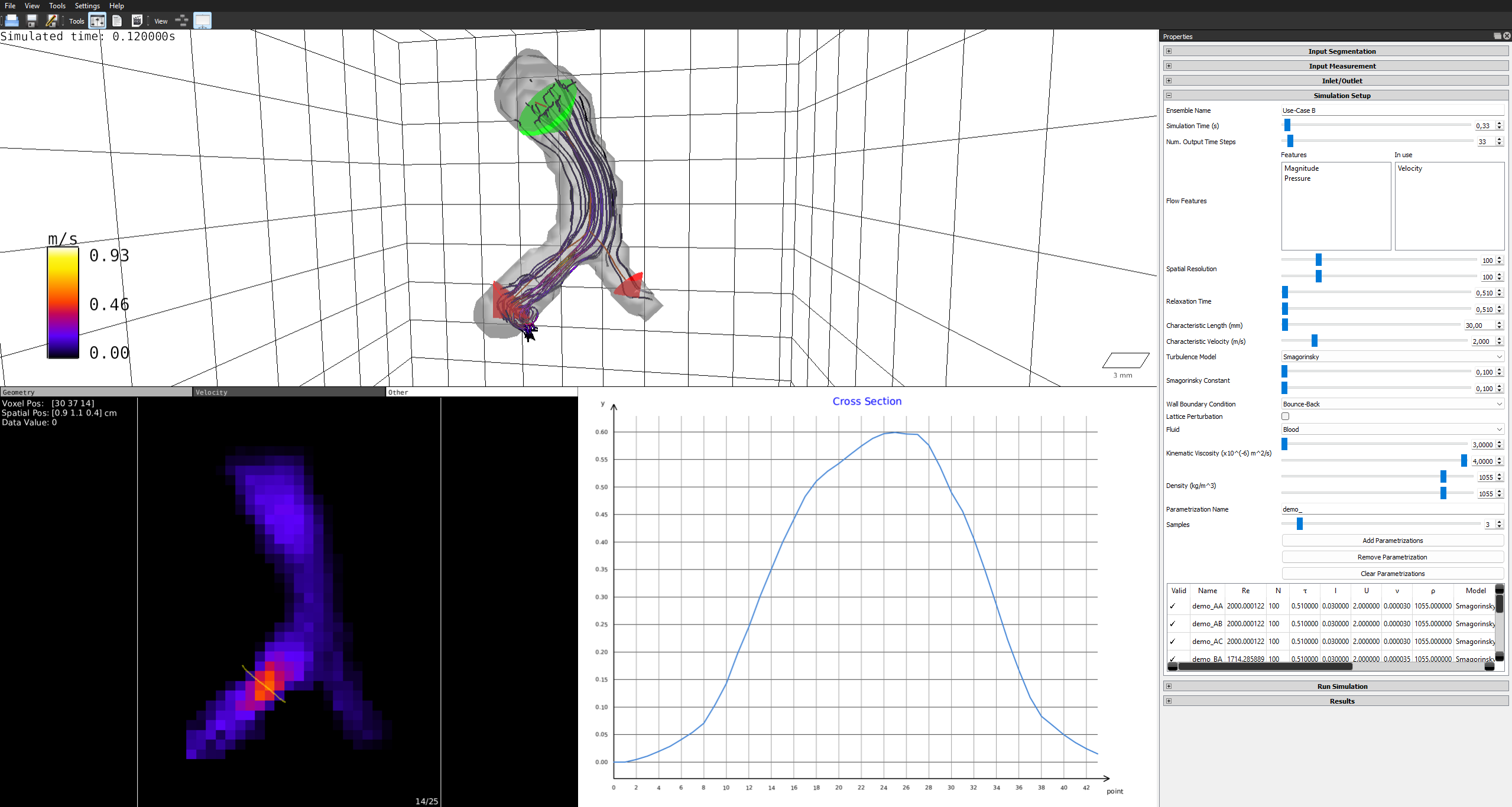}
    \caption{Simulation result for Use Case B similar to \autoref{fig:uc-a}.}
    \label{fig:uc-b_simulation}
\end{figure}

Cardiovascular imaging researchers working with 4D-PC-MRI data obtain time series of 3D velocity fields. Here, $12$ velocity maps of a rat’s pulmonary artery, sampled at $10$ \si{\milli\second} intervals, were stored in .vti format \cite{VTI-explanation}. 
The goal is to reproduce these measurements with simulations: while MRI data are noisy and low resolution, simulations can provide noise-free, high-resolution fields but require modeling assumptions. To account for these, we generate simulation ensembles with varying parameters and compare results in Use Case C.
In Use Case B, the measured velocity is used as inlet boundary condition, and the simulation evolves from there. Two workflows were developed to preprocess such data and perform simulations, involving the following steps:

\begin{itemize}
\item For preprocessing, open the workspace \textbf{use\_case\_B\_segmentation.vws} and select the measured data using the \textit{Input} panel. The \textit{Data Settings} panel allows to correct the data format and enable aggregation. Since the segmentation works best with high-contrast images, a single time step might not be suited to obtain the segmentation of the desired vessel's lumen. Depending on the imaging method used, parts of it may only be bright in a certain time step, other parts in a different time step, hence, we allow to aggregate along the temporal axis, e.g., using the maximum, mean voxel value, and the sum. The \textit{2D Rendering} panel, allows to adjust the transfer function. 
\item Using the mouse, foreground and background seeds can be placed interactively by the user which are fed into the random-walker segmentation algorithm~\cite{random-walker}. In our example, we choose to only segment the pulmonary trunk contained in the measurement. 
\item In the \textit{Output} panel, the segmentation mask can be stored in various file formats, such as .vti. 
\item Next, open the workspace \textbf{use\_case\_B\_simulation.vws} and upload the created segmentation along with the measured velocity maps. 
\item Proceed with configuring parameters, running the simulation, and saving the results, similar to Use Case A within \textit{Inlet/Outlet}, \textit{Simulation Setup}, \textit{Run Simulations} and \textit{Results}. In contrast to Use Case A, we use the measurement as initialization for the velocity inlet boundary condition and run an ensemble, with three members.
\end{itemize}
\autoref{fig:uc-b_segmentation} shows a sample result for the segmentation and \autoref{fig:uc-b_simulation} displays one of the calculated simulations.

\subsection{Use Case C: Analysis of simulation ensembles}\label{uc: uc-c}
\begin{figure}[ht]
    \centering
    \includegraphics[width=\columnwidth]{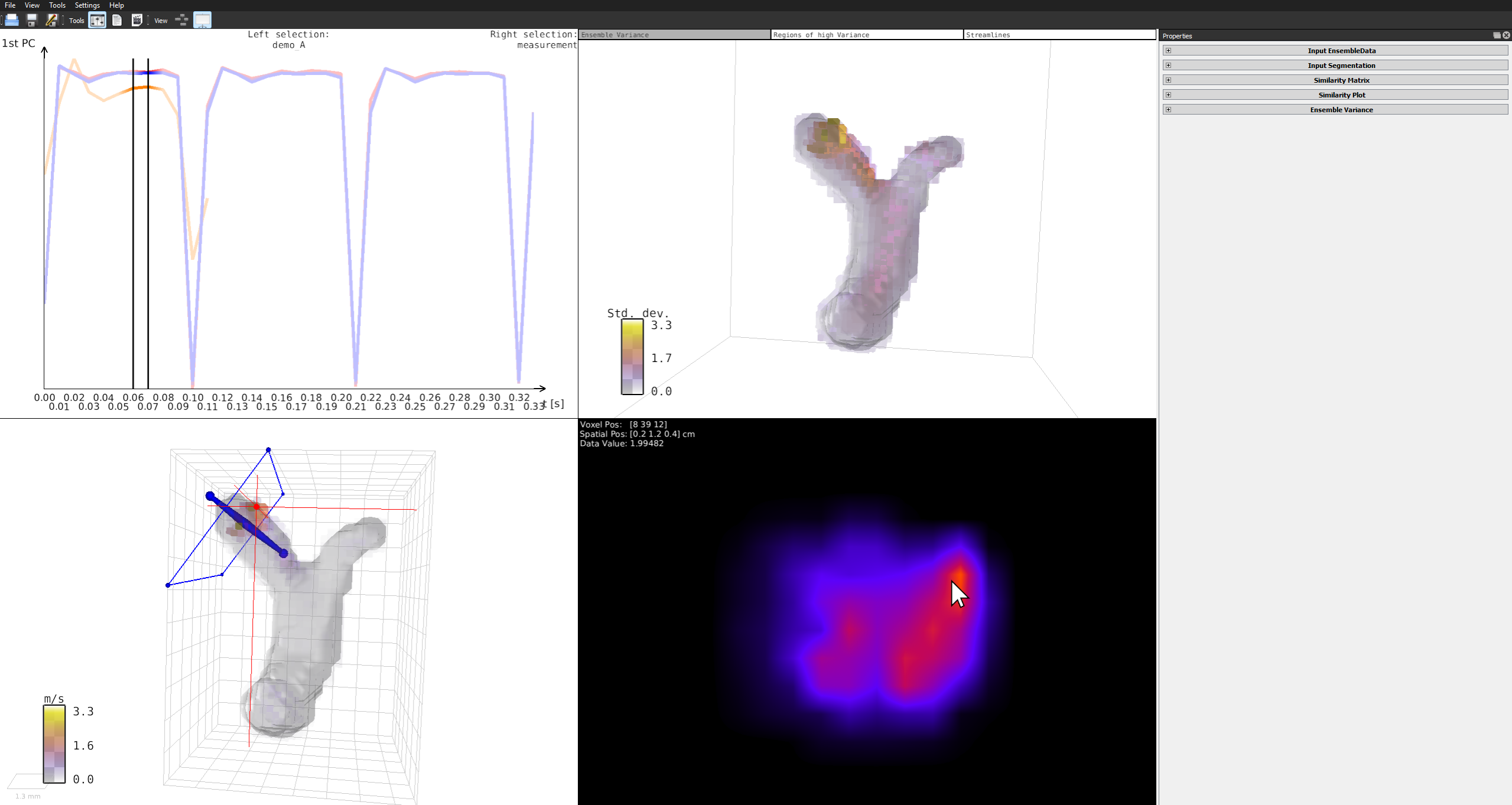}
    \caption{Multiple linked views visualize the 2D Similarity Plot (top left), deviation of two members (top right), the raycasted (bottom left) and slice-rendered voxel-wise difference (bottom right). The user can change the selected members in the similarity plot. Tabs in the top right view switch between member deviation, connected components and streamlines.}
    \label{fig:uc-c}
\end{figure}

In both domains, analysis depends on the task at hand. Typical questions include assessing the effect of specific parameters (e.g., velocity magnitude) or comparing simulations with real-world measurements to identify the best-fitting parameters. Similarity measures for this purpose have been presented in \cite{leistikow2020interactive}. To apply them, researchers can perform the following steps:
\begin{itemize}
    \item Open the workspace \textbf{use\_case\_C.vws} and select the simulation ensemble that has been created with either Use Case A or B using the \textit{Input} panel. 
    \item Next, the \textit{Similarity Plot} and \textit{Ensemble Variance}, may be used to obtain an overview over the ensemble. Two members, e.g., the measurement and a simulation similar to it, can be selected from the Similarity Plots and compared directly by visualizing their voxel-wise difference. It is represented both as raycasting and as slice-rendering for reading off exact voxel values. 
    The similarity plot is obtained through a vector-field similarity measure proposed by Leistikow et al.~\cite{leistikow2020interactive}. It is evaluated for each pair of time steps of each simulation or measurement and stored in a similarity matrix. This matrix is then projected into a single dimension using multi-dimension scaling (MDS) and plotted over time to obtain a 2D Similarity Plot where the vertical distance between runs represents their dissimilarity for a selected point in time. For details, we refer to their publication.
\end{itemize}
\autoref{fig:uc-c} shows the ensemble visualization with our configuration settings.

\section{Discussion}\label{sec:discussion}
To evaluate our software, we conducted two workshop sessions with three researchers. Session S1 involved a master’s student in Fluid-Structure Interaction (P1) and a doctoral student in CFD (P2), while Session S2 included a postdoctoral researcher in cardiovascular imaging (P3). Each session covered the full data pipeline (\autoref{fig:ensemble_pipeline}), with S1 focusing on Use Cases A (\ref{uc: uc-a}) and C (\ref{uc: uc-c}), and S2 on Use Cases B (\ref{uc: uc-b}) and C. Each Session ended with discussions on the comparison factors (\autoref{tab:comparison_factors}), the system’s utility for participants’ work, and potential improvements. We summarize their feedback below.

\subsection{CFD Domain}\label{feedback:S1}
Participants P1 and P2 saw potential in Use Case A, especially the rapid configuration of inlet and outlet boundary conditions via Voreen’s GUI. It accelerates setup and supports collaboration with non-coding researchers. They also valued the intermediate visualization of computational steps for analyzing results and quality.

For experienced users automatic boundary placement at vessel endpoints was seen as too restrictive. Both suggested enabling variable boundary positioning within the GUI, with P2 stressing this would be a major improvement given the time-consuming nature of such tasks.

Regarding Use Case C, both participants considered it irrelevant for their workflows. CFD research typically relies on standard benchmarks, and Use Case C was is aligned with these.

\subsection{Cardiovascular Imaging Domain}\label{feedback:S2}
P3 recognized the value of the integrated data pipeline, highlighting the performance of image processing, simulation configuration, and result analysis within a single software environment. Similar to P1 and P2, P3 also noted the advantages of the GUI-based boundary condition configuration, which facilitates efficient creation of simulation ensembles.

From P3’s perspective, Use Case C contributed to simulation evaluation by enabling direct visual comparison between measured and simulated data. For instance, we created a simulation intended to replicate flow measurements in a rat's pulmonary artery. Initial results appeared promising, however, a critical vortex structure observed in the measurement was absent in the simulation, leading to a classification of the simulation as inadequate.

P3, an experienced SimVascular and Voreen user, compared our system to it during Session S2. He noted that both platforms support volume rendering, but SimVascular provides different simulation configuration, such as local mesh-size adaptation and boundary condition refinement. While Use Case C offered useful visualizations, P3 found them limited to specific tasks. For the variety of fluid mechanics conditions in human body, such as bifurcations, branching, curvatures and disease specific flow profiles, we need more predefined workspaces for users to interact with. Adapting these workflows requires learning Voreen, which he considered more challenging than SimVascular and ParaView. In addition, he perceived SimVascular and ParaView as offering a broader range of usage.

\section{Conclusion and Future Work}\label{sec:conclusion}
We presented a novel open source software system for the generation and analysis of flow simulation ensembles. In particular, we examined use cases for typical simulation and analysis tasks. Image preprocessing, simulation configuration and simulation analysis are supported by a GUI allowing the users to interactively setup, adjust, and visualize measured data, geometric volume files, and simulation results. By integrating all steps into one software solution, we provide a comprehensive tool, especially for researchers without coding experience and newcomers in the field. Our system supports different groups of researchers in medicine, biology, and numerics interested in exploratory analysis of medical and volumetric flow data. 

In future work, we plan to integrate more customizable boundary condition placement, allowing users to configure 
$x$, $y$, and $z$- positions along the mesh and shift boundary elements via drag-and-drop (\autoref{feedback:S1}). Session S2 feedback (\autoref{feedback:S2}) further suggested adding MRI-specific configurations (e.g., deformable vessel walls as in SimVascular) and incorporating human data into predefined workflows to enhance clinical relevance.


\section*{CRediT authorship contribution statement}
\textbf{Simon Leistikow:} Writing - review \& editing, Conceptualization, Validation, Methodology, Software\\
\textbf{Thomas Miro:} Writing - original draft \& review \& editing, Conceptualization, Validation, Formal analysis, Visualization\\
\textbf{Adrian Kummerländer:} Writing - review \& editing, Methodology, Software, Validation\\
\textbf{Ali Nahardani:} Writing - review\& editing, Validation, Resources\\
\textbf{Katja Grün:} Writing - review \& editing, Resources\\
\textbf{Markus Franz:} Writing - review \& editing, Resources\\
\textbf{Verena Hoerr:} Writing - review \& editing, Supervision, Resources\\
\textbf{Mathias J. Krause:} Writing - review \& editing, Supervision\\
\textbf{Lars Linsen:} Writing - review \& editing, Supervision\\

\section*{Declaration of Competing Interest}
The authors declare that they have no known competing financial interests or personal relationships that could have appeared to influence the work reported in this paper.

\section*{Declaration of generative AI and AI-assisted technologies in the writing process}
During the preparation of this work the authors used ChatGPT and a generative AI hosted by the University of Münster in order to improve the readability and language of the manuscript. After using this tool/service, the author(s) reviewed and edited the content as needed and takes full responsibility for the content of the published article.

\section*{Funding}
This work was supported by the Deutsche Forschungsgemeinschaft (DFG) under project number 468824876.

\section*{Ethical Approval}
Not applicable.



\bibliographystyle{elsarticle-num} 

\input{main.bbl}





\end{document}

%% file: macro.tex
\usepackage{framed, color} 
\usepackage{xcolor} 
\usepackage{parskip} 

\usepackage{graphicx}
\usepackage{amsmath}
\usepackage{amssymb}
\usepackage{booktabs}
\usepackage{url}
\usepackage{hyperref}
\usepackage{balance}
\usepackage{times}
\usepackage{float}
\usepackage{comment}
\usepackage{xcolor}
\usepackage{wasysym}
\usepackage{rotating}
\usepackage{dsfont}
\usepackage{tikz}
\usepackage{siunitx}
\usepackage{svg}

%% file: figures/d3q19.tikz
\begin{tikzpicture}[>=latex]
 \centering
    \draw[densely dashed] (-2,0,0) -- (0,0,0);
    \draw[densely dashed] (-2,0,0) -- (-2,2,0);
    \draw[densely dashed] (-2,0,0) -- (-2,0,2);
    \draw[densely dashed] (-2,0,2) -- (0,0,2);
    \draw[densely dashed] (-2,2,0) -- (0,2,0);
    \draw[densely dashed] (-2,0,2) -- (-2,2,2);
    \draw[densely dashed] (-2,2,0) -- (-2,2,2);
    \draw[densely dashed] (-2,2,2) -- (0,2,2);
    \draw[densely dashed] (0,0,0) -- (0,0,2);
    \draw[densely dashed] (0,0,0) -- (0,2,0);
    \draw[densely dashed] (0,0,2) -- (0,2,2);
    \draw[densely dashed] (0,2,2) -- (0,2,0);
 	\draw[densely dashed] (-2,0,1) -- (0,0,1); 
	\draw[densely dashed] (-2,0,1) -- (-2,2,1); 
	\draw[densely dashed] (-2,2,1) -- (0,2,1); 
	\draw[densely dashed] (0,2,1) -- (0,0,1); 
	\draw[densely dashed] (-1,0,0) -- (-1,0,2); 
	\draw[densely dashed] (-1,0,2) -- (-1,2,2); 
	\draw[densely dashed] (-1,2,2) -- (-1,2,0); 
	\draw[densely dashed] (-1,2,0) -- (-1,0,0); 
	\draw[densely dashed] (-2,1,0) -- (0,1,0); 
	\draw[densely dashed] (0,1,0) -- (0,1,2); 
	\draw[densely dashed] (0,1,2) -- (-2,1,2); 
	\draw[densely dashed] (-2,1,2) -- (-2,1,0);   
    \draw[->,thick](-1,1,1) -- (-1,1,0);
    \draw[->,thick](-1,1,1) -- (-2,1,0);
    \draw[->,thick](-1,1,1) -- (-1,2,0);
    \draw[->,thick](-1,1,1) -- (-1,0,0);
    \draw[->,thick](-1,1,1) -- (0,1,0);
    \draw[thick,fill=cyan](-1,1,0) circle(2pt);     
	\draw[thick,fill=green](-2,1,0) circle(1.5pt);
  	\draw[thick,fill=green](-1,2,0) circle(1.5pt);
  	\draw[thick,fill=green](-1,0,0) circle(1.5pt);
	\draw[thick,fill=green](0,1,0) circle(1.5pt);
    \draw[->,thick](-1,1,1) -- (-1,2,1);
    \draw[->,thick](-1,1,1) -- (0,1,1);
    \draw[->,thick](-1,1,1) -- (-2,1,1);
    \draw[->,thick](-1,1,1) -- (-1,0,1);
    \draw[->,thick](-1,1,1) -- (-2,0,1);
    \draw[->,thick](-1,1,1) -- (0,0,1);
    \draw[->,thick](-1,1,1) -- (-2,2,1);
    \draw[->,thick](-1,1,1) -- (0,2,1);
	\draw[thick,fill=cyan](-1,2,1) circle(2pt);
    \draw[thick,fill=cyan](0,1,1) circle(2pt);
    \draw[thick,fill=cyan](-1,0,1) circle(2pt);
    \draw[thick,fill=cyan](-2,1,1) circle(2pt);
    \draw[thick,fill=green](-2,0,1) circle(1.5pt);     
  	\draw[thick,fill=green](0,0,1) circle(1.5pt);
    \draw[thick,fill=green](-2,2,1) circle(1.5pt);     
  	\draw[thick,fill=green](0,2,1) circle(1.5pt);
	\draw[->,thick](-1,1,1) -- (-1,1,2);
    \draw[->,thick](-1,1,1) -- (-1,0,2);
	\draw[->,thick](-1,1,1) -- (-2,1,2);
	\draw[->,thick](-1,1,1) -- (0,1,2);
	\draw[->,thick](-1,1,1) -- (-1,2,2);
    \draw[thick,fill=cyan](-1,1,2) circle(2pt);
	\draw[thick,fill=green](-1,0,2) circle(1.5pt);
	\draw[thick,fill=green](-2,1,2) circle(1.5pt);
	\draw[thick,fill=green](0,1,2) circle(1.5pt);
	\draw[thick,fill=green](-1,2,2) circle(1.5pt);
    \draw[thick,fill=orange](-1,1,1) circle(3pt);
  \end{tikzpicture}
\hspace{2em}
\begin{tikzpicture}
	\draw[->,thick] (0,0,0) -- (.5,0,0);
    \node[anchor=north] at (.5,0,0) {$x$};
    \draw[->,thick] (0,0,0) -- (0,.5,0);
    \node[anchor=east] at (0,.5,0) {$y$};
	\draw[->,thick] (0,0,0) -- (0,0,.5);
	\node[anchor=north west] at (0,0,.5) {$z$};
\end{tikzpicture}